\documentclass[pra, aps, twocolumn, groupedaddress, showpacs, superscriptaddress]{revtex4}

\usepackage{slashed}
\usepackage{graphicx}
\usepackage{subfigure}
\usepackage[usenames, dvipsnames]{color}
\usepackage{graphics}
\usepackage{hyperref}
\usepackage{bm}
\usepackage{amsfonts}
\hypersetup{backref,
colorlinks=true,
urlcolor=blue,
linkcolor=blue,
linktoc=page,
citecolor=blue}

\everymath{\displaystyle} 
\begin{document}

\title{Entanglement sudden death and revival in quantum dark-soliton qubits}

\author{Muzzamal I. Shaukat}
\affiliation{CeFEMA, Instituto Superior T\'ecnico, Universidade de Lisboa, Lisboa, Portugal}
\affiliation{University of Engineering and Technology, Lahore (RCET Campus), Pakistan}
\email{muzzamalshaukat@gmail.com}
\author{Eduardo. V. Castro}
\affiliation{CeFEMA, Instituto Superior T\'ecnico, Universidade de Lisboa, Lisboa, Portugal}
\author{Hugo Ter\c{c}as}
\affiliation{Instituto de Plasmas e Fus\~ao Nuclear, Lisboa, Portugal}
\email{hugo.tercas@tecnico.ulisboa.pt}

\pacs{67.85.Hj 42.50.Lc 42.50.-p 42.50.Md }

\begin{abstract}
We study the finite time entanglement dynamics between two dark soliton qubits due to quantum fluctuations in a quasi one dimensional Bose-Einstein condensates. Recently, dark solitons are proved to be an appealing platform for qubits due to their appreciably long life time. We explore the entanglement decay for an entangled state of two phonon coherences and the qubits to be in the diagonal basis of so called Dicke states. We observe the collapse and revival of the entanglement, depending critically on the collective damping term but independent of the qubit-qubit interaction for both the initial states. The collective behavior of the dark soliton qubits demonstrate the dependence of entanglement evolution on the interatomic distance.
 
 
\end{abstract}

\maketitle

\section{Introduction}
Two basic features to distinguish the classical world from the quantum world are the superposition and entanglement, being the cornerstone of the rapidly developing field of quantum information and computation \cite{Nielsen2000}. In recent years, the interaction between qubits and the environment has attracted a great deal of attention motivated by the possibility of controlling and exploiting the entanglement dynamics \cite{Malinovsky2004, Plenio1999,  Serafini2006, Ficek2002}. The occurence of spontantaneous emission due to the coupling of qubits with the reservoir, leading to the irreversible loss of quantum information therein encoded, has been regarded as the main obstacle in practical usage of entanglement. \par

Much work has been done to understand the decoherence dynamics for a pair of qubits interacting with different reservoirs \cite{Ficek2003,Ficek2004,Ficek2003a}. As it is currently known, two spatially separated qubits, initially prepared in a product of two pure states, get entangled as time evolves, producing to the creation of a so-called {\it transient entanglement} in the system \cite{Ficek2003,Ficek2004}. Conversely, Yu and Eberly discovered a finite-time disentanglement of two initially-entangled qubits in contact with pure dissipative environments \cite{Eberly2004,Eberly2006}. This effect is currently known under the name of {\it entanglement sudden death}, and has been confirmed with experiments performed both with photonic \cite{Almeida2007} and atomic systems \cite{Laurat2007}. \par

In this context, Bose-Einstein condensates (BECs) have attracted a great deal of interest during the last decades, since the macroscopic character of the wavefunction allows BECs to display pure-state entanglement at macroscopic scales \cite{Davis1995,Anderson1995,Parkins1998}. A scheme to generate entangled states in a bimodal BEC has been firstly announced in Refs. \cite{Hines2003, Micheli2003}; the investigation of the macroscopic superposition based on matter waves has been achieved with BEC Josephson junctions \cite{Gordon1999, Cirac1998, Ruostekoskil1998}. Moreover, light scattering with BECs have been used to enhance their nonlinear properties in superradiance experiments \cite{Inouye1999}, and to show the possibility of matter wave amplification \cite{Kozuma1999} and nonlinear wave mixing \cite{Deng1999}. Entanglement dynamics for coupled BECs have been investigated in \cite{Sanz2003}. 

Another important manifestation of the macroscopic nature of BEC is the dark soliton (DS), a structure resulting from the detailed balance between the dispersive and nonlinear effects, appearing also in other physical systems \cite{kivshar, Denschlag, burger}. The dynamics and stability of DSs in BECs have been a subject of intense research over the last decade \cite{Dziarmaga2003,Jackson2007}.  The dynamical evolution of DS entanglement and how its stability is affected by quantum fluctuations has been studied in Ref. \cite{Mishmash2009}. Collision-induced entanglement between fast moving matter-wave solitons using the Born approximation has been studied in BECs displaying attractive interactions \cite{Lewenstein2009}. Moreover, the study of collective aspects of soliton gases \cite{gael2005} bring DSs towards applications in many body physics \cite{tercas}. In a recent publication, we have shown that DSs trapping an impurity can behave as qubits in quasi-1D BECs \cite{muzzamal2017}, being excellent candidates to store information as consequence of their appreciably long lifetimes ($\sim 0.01-1$s). DSs qubits thus offer an appealing alternative to quantum optical of solid-state platforms, as information processing involves only phononic degrees of freedom: the quantum excitations on top of the BEC state.\par

Ghasemian et al. \cite{Ghasemian2016} described the collapse-revival phenomenon, after studying the dynamics of BEC atoms interacting with a single-mode laser field. The possibility of generating entangled Schr\"odinger cat states by using a BEC trapped in a double-well potential has been investigated \cite{Fisher1989,Huang2006}. In these works, it is reported that the revival of the initial state can be used as an unambiguous signature of the coherent macroscopic superposition, as opposed to an incoherent mixture. Also, two-impurity qubits surrounded by a BEC reveal the influence of quantum reservoirs on effects such as sudden death, revival, trapping and generation of entanglement \citep{McEndoo2013}.

In this paper, we investigate the dynamics of the entanglement produced between two dark-soliton qubits in a quasi-one dimensional Bose-Einstein condensate. As described in Ref \cite{muzzamal2017}, the qubits are produced by trapping impurities inside the potential created by the dark solitons. Moreover, the phonons (quantum fluctuations on top of the background density) play the role of a quantum reservoir. We show the occurrence of entanglement sudden death by computing the time evolution of the Wootters' concurrence and showing that it vanishes for a finite time. We further demonstrate that the concurrence dynamics critically depends on the distance between the DS qubits. 		\par

The paper is organized as follows: In sec. II, we model the mean-field dynamics of the BEC and the impurities by using the Gross-Pitaevskii (GP) and Schr\"odinger equations, respectively. We compute the coupling between the DS-qubits and the phonons. In sec. III,  we describe the Markovian master equation and extract the density matrix elements for the collective DS qubit states. The use of concurrence as a measure of the entanglement is discussed in sec. IV. Finally, a summary and discussion of the investigation is presented in Sec. V.
\begin{figure}[t!]
\includegraphics[scale=0.35]{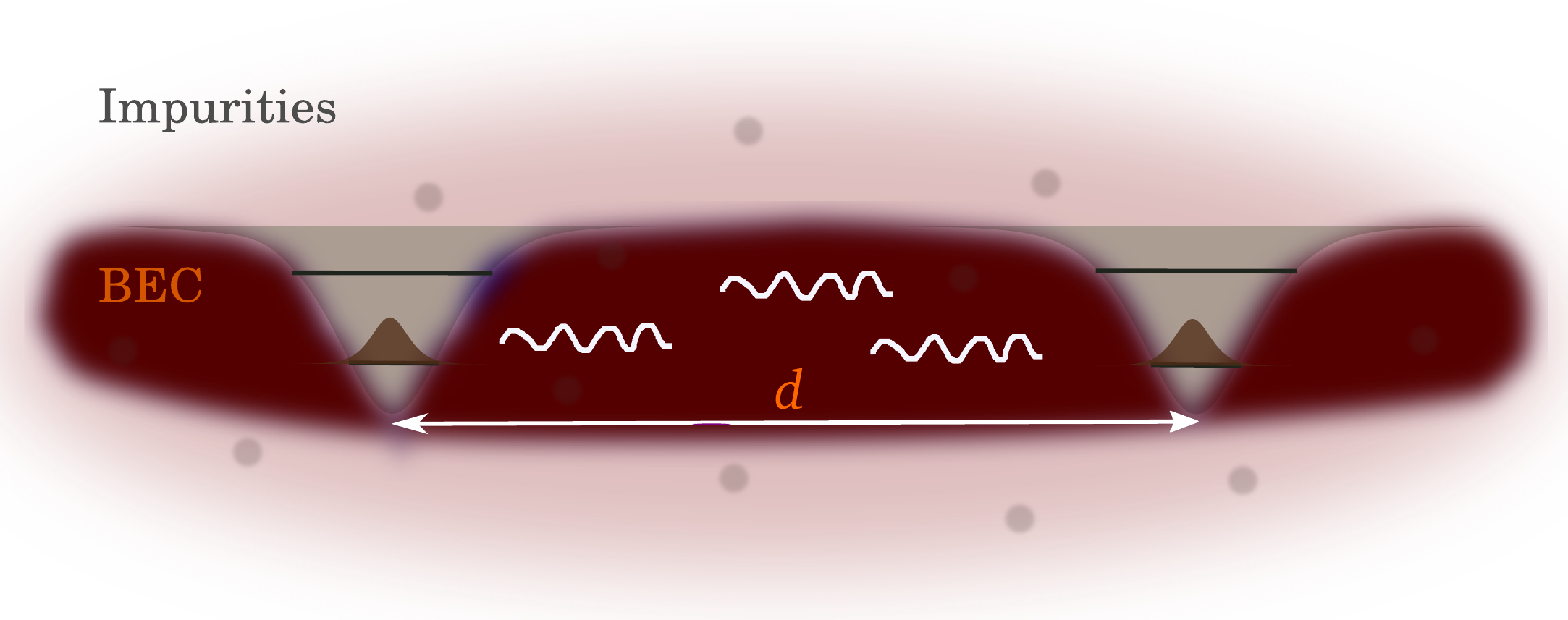}
\caption{(color online) Schematic representation of two dark-soliton qubits in a cigar shaped quasi-1D Bose-Einstein condensates immersed in a dilute gas of impurities. The localized depressions in the density represent the dark solitons, while the wiggly lines represent the phonons, i.e. the elementary excitations composing the quantum reservoir.}
\label{fig_scheme}
\end{figure}

\begin{figure}[t!]
\includegraphics[scale=0.65]{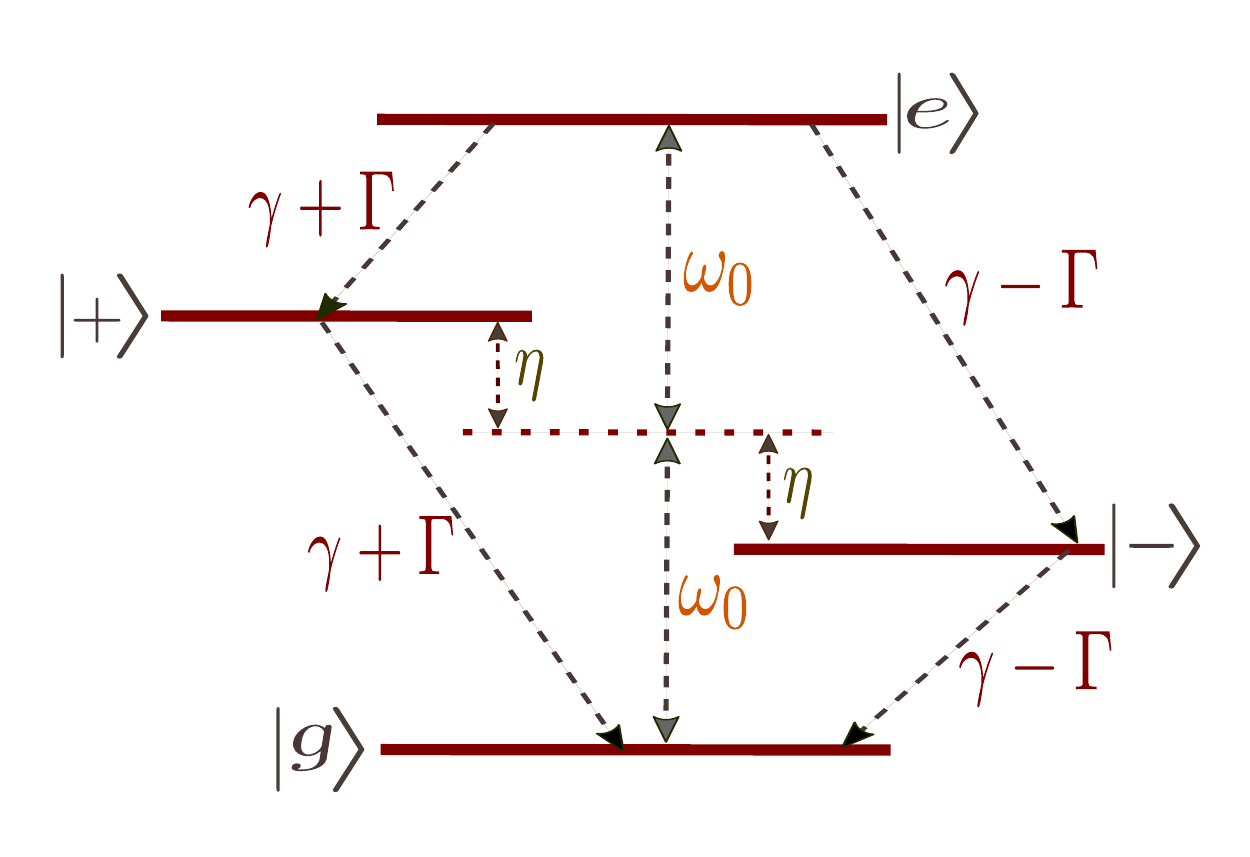}
\caption{(color online) Collective states of two dark-soliton qubits.}
\label{fig_Collective States}
\end{figure}

\section{Theoretical Model and Quantum Fluctuations}

At the mean field level, the system is governed by Gross-Pitaevskii (GP) equations,
\begin{eqnarray}
i\hbar \frac{\partial \Psi}{\partial t}&=&-\frac{\hbar ^{2}}{2m_1}\frac{
\partial^{2} \Psi}{\partial x^{2}}+g\left\vert \Psi\right\vert ^{2}\Psi
+\chi\left\vert \Phi\right\vert ^{2}\Psi, \label{gp1} \\
i\hbar \frac{\partial \Phi}{\partial t}&=&-\frac{\hbar ^{2}}{2m_2}\frac{
\partial^{2} \Phi}{\partial x^{2}}+\chi\left\vert \Psi \right\vert ^{2}\Phi,  \label{gp2}
\end{eqnarray}
where, $\chi$ is the BEC-impurity coupling constant, $g$ is the BEC particle-particle interaction strength, and $m_1$ and $m_2$ denote the BEC particle and impurity masses, respectively. Here, the discussion is restricted to repulsive interacions ($g>0$) where the dark solitons are assumed to be not disturbed by the presence of impurities. To achieve this, the impurity gas is considered to be much less massive than the BEC particles and sufficiently dilute, i.e. $\vert \Psi\vert^2 \gg \vert\Phi\vert^2 $ (experimental realization can be found in \cite{muzzamal2017}). Therefore, the soliton behave like a potential for the impurities (considered to be free particles) i.e.,
\begin{equation}
i\hbar \frac{\partial \Phi}{\partial t}=-\frac{\hbar ^{2}}{2m_2}\frac{
\partial^{2} \Phi}{\partial x^{2}}+\chi\left\vert \psi _{\rm sol}\right\vert ^{2}\Phi,  \label{sch. eq. without soliton}
\end{equation}
where a singular nonlinear solution corresponds to the $i$th ($i=1,2$) soliton profile is given by \cite{zakharov72, huang} 

\begin{equation}
\psi^{(i)}_{\rm sol}(x)=\sqrt{n_{0}}\tanh \left[ (x -x_i)/\xi \right],
\end{equation}
where $x_{i}=\pm d/2$ determines the position of the solitons, $\xi =\hbar /\sqrt{mn_{0}g_{11}}$ is the healing length and $n_0\sim 10^{8}-10^{9} m^{-1}$ is the typical linear density. For well separated solitons, $ d\gg \xi$, the internal level structure of the two qubits is assumed to be equal. Each qubit is characterized by its ground $g_{i}$ and excited levels $e_{i}$ ($i=1,2$), separated by a gap frequency $\omega _{0}=\hbar(2\nu -1)/(2m\xi ^{2})$ (see Fig. (\ref{fig_scheme})). Here, 
\begin{equation}
\nu=-1+\sqrt{1+4\chi/g}
\end{equation}
is a parameter controlling the number of bound states created by the DS, which operates as a qubit (two-level system) in the range $0.33\leq\nu <0.80$ \cite{muzzamal2017}. Typical experimentally accessible conditions provide longitudinal and transverse trap frequencies of $\omega _{z}/2\pi\sim (15-730)$ Hz $\ll\omega_{r}/2\pi= (1-5)$ kHz, and the corresponding length $l_{z}=(0.6-3.9)$ $\mu$m \cite{parker2004}. More recent experiments lead to much less pronounced trap inhomogeneities by creating much larger traps, $l_z\sim 100$ $\mu$m \cite{schmiedmayer2010}. 

\subsection{Quantum fluctuations} 

The total BEC quantum field includes the DSs wave functions and quantum fluctuations, $\Psi_{i}(x)=\psi^{(i)}_{\rm sol}(x)+\delta \psi_{i}(x)$, where 
 \begin{equation}
\delta \psi_{i}(x)=\sum_k \left(u^{(i)}_k(x) b_k +v^{*(i)}_k(x)b^{\dagger}_k \right).
\label{phonon eigen function}
\end{equation}
Here, $b_k$ are the bosonic operators verifying the commutation relation $[b_{k},b^{\dagger}_{q}]=\delta_{k,q}$. The amplitudes $u_k(x)$ and $v_k(x)$ satisfy the normalization condition $\vert u_k(x)\vert ^2 -\vert v_k(x)\vert ^2=1$ and are explicitly given by \cite{Martinez2011} 
\begin{eqnarray*}
&&\hspace{-0.5cm}\left. u^{(i)}_{k}(x)= e^{ik(x-x_{i})}\sqrt{\frac{1}{4\pi \xi}}\frac{\mu}{\epsilon_{k}}%
\right. \times \\
&& \hspace{-0.6cm}\left. \left[\left((k\xi)^{2}+\frac{2\epsilon_{k}}{\mu}\right)\left(\frac{k\xi}{2}+i\tanh\left(\frac{x-x_{i}}{\xi}\right)\right)+\frac{k\xi}{%
\cosh^{2}\left(\frac{x-x_{i}}{\xi}\right)}\right] \right.,
\end{eqnarray*}%
\begin{eqnarray*}
&&\hspace{-0.5cm}\left. v^{(i)}_{k}(x)= e^{-ik(x-x_{i})}\sqrt{\frac{1}{4\pi \xi}}\frac{\mu}{\epsilon_{k}}%
\right. \times \\
&& \hspace{-0.6cm} \left. \left[\left((k\xi)^{2}-\frac{2\epsilon_{k}}{\mu}\right)\left(\frac{k\xi}{2}+i\tanh\left(\frac{x-x_{i}}{\xi}\right)\right)+\frac{k\xi}{%
\cosh^{2}\left(\frac{x-x_{i}}{\xi}\right)}\right] \right..
\end{eqnarray*}%
The total Hamiltonian then reads
\begin{equation}
H=H_{\rm q}+H_{\rm p}+H_{\rm int}.
\label{Tot Ham}
\end{equation}
The term $H_{\rm q}$ describes the dark-solitons (qubits) Hamiltonian, which is given by
\begin{equation}
H_{\rm q}=\sum^{2}_{i=1}\hbar \omega _{0}\sigma _{z} ^{(i)} \label{qubit ham.}
\label{system Hamilt.}
\end{equation}
with $\sigma_z^{(i)}=a_1^{(i)\dagger} a_1^{(i)}- a_0^{(i)\dagger} a_0^{(i)}$ being the effective spin operator of the respective qubit. The phonon (reservoir) Hamiltonian is given by
\begin{equation}
H_{\rm p }=\sum_k \epsilon _{k}b_{k}^{\dagger} b _{k}, \label{phonon ham.}
\label{envir. hamilt.}
\end{equation}
with the Bogoliubov spectrum $\epsilon _{k}=\mu \xi \sqrt{k^{2}(\xi^{2}k^{2}+2)}$ and the chemical potential $\mu=gn_{0}$.  Finally, the interaction Hamiltonian $H_{\rm int}$ can be expliclity written as 
\begin{equation}
H_{\rm int}=\sum_{i,j}\chi\int dx\Phi _{j}^{\dag }\Psi _{i}^{\dag }\Psi _{i}\Phi _{j},
\label{Int. Ham.}
\end{equation}
where $\Phi_{j}(x)$ describes the impurity wave function in the presence of DS potential and spannable interms of bosonic operators $a_{l}$   
\begin{equation}
\Phi_{j}(x)=\sum_{n=0}^1 \phi^{(j)}_{n}(x) a^{(j)}_{n},
\label{eq_dec2}
\end{equation}
with the ground state $\phi_0(x)={\rm sech} \left[ (x -x_i)/\xi \right]/\sqrt{2\xi}$ and the excited state $\phi_1(x)=i \sqrt{3}\tanh \left[ (x -x_i)/\xi \right]\phi_0(x)$.
Therefore, Eq. (\ref{Int. Ham.}) can be decomposed as
\begin{equation}
H_{\rm int}=H_{\rm int}^{(0)}+H_{\rm int}^{(1)}+H_{\rm int}^{(2)}, 
\label{Inter. hamilt.}
\end{equation}
containing zeroth, first and second order terms in the bosonic operators $b_k$ and $b_k^\dagger$. The higher-order term $\sim \mathcal{O}(b_{k}^{2})$ is ignored, in consistency with the Bogoliubov approximation performed in Eq. (\ref{phonon eigen function}) owing to the small depletion of the condensate. This approximation is well justified in the case of two-level systems, as the inexistent higher excited states cannot be populated via two-phonon processes. The first part of Eq. (\ref{Inter. hamilt.}) corresponds to 
\begin{equation}
H_{\rm int}^{(0)}=n_{0}\chi\sum_{i=1}^{2}\sum_{n,n^{'}=0}^{1}a_{n}^{\dag (i) }a_{n^{'}}^{(i)}f_{n,n^{'}}^{(i)} .
\end{equation}
with $f^{(i)}_{n,n^{'}}=\int dx\phi _{n}^{\dag (i) }(x) \phi_{n^{'}}^{(i)}(x) \rm tanh^{2}\left[(x -x_i)/\xi\right]$, which can be omitted by renormalizing the qubit frequency $\tilde\omega_{0}\approx\omega_{0}+n_{0}\chi$. The first order term is given by
\begin{eqnarray}
H_{\rm int}^{(1)} &=&\sum_{k,i=1}^{2}\sum_{n,n^{'}=0}^{1}a_{n}^{(i)\dag }a^{(i)}_{n^{'}}\left[ b_{k}g_{n,n^{'}}^{(i)}(k)\right.	\nonumber \\&+& \left. b_{k}^{\dag}g_{n,n^{'}}^{(i)\ast}(k)\right] \label{Final Hmailtonian 1}
\end{eqnarray}%
where
\begin{eqnarray*}
g_{n,n^{'}}^{i,j}(k) = \sqrt{n_{0}}\chi\int dx\phi^{(j) \dag} _{n}(x) \phi^{(j)}_{n^{'}}(x) \tanh\left(\frac{x -x_i}{\xi }\right) u^{(i)}_{k}.
\end{eqnarray*} 
Eq. (\ref{Final Hmailtonian 1}) contains both interband ($n \neq n'$) and intraband ($n = n'$) terms. However, within the RWA, the qubit transition can only be driven by near-resonant phonons, for which the intraband terms $\vert g_{00}^{(i)}(k) \vert$ and $\vert g_{11}^{(i)}(k)\vert$ are much smaller than the interband term  $\vert g_{01}^{(i)}(k)\vert=\vert g_{10}^{(i)}(k)^*\vert$. As such, we obtain 
\begin{eqnarray*}
H_{\rm int}^{(1)} &=&\sum_{k,i=1}^{2}\left( g^{(i)}(k)\sigma^{(i)} _{+}b_{k}+ g^{(i) \ast}(k) \sigma^{(i)} _{-}b_{k}^{\dag }\right)+{\rm h.c.} .
\end{eqnarray*}
Here, $g^{(i)}(k)=g_{n,n^{'}}^{i,j}(k)$, $\sigma^{(i)} _{+}=a_{1}^{(i)\dag }a^{(i)}_{0}$ and vice versa. The counter-rotating terms proportional to $b_k \sigma_-$ and $b_k^\dagger\sigma_+$ that do not conserve the total number of excitations are dropped by invoking the RWA. The accuracy of such an approximation can be verified in Ref. \citep{muzzamal2017}, where it is shown that the emission rate $\gamma$ is much smaller than the qubit transition frequency $\omega_0$ for DS qubits.
 
\section{Master Equation}

We derive the master equation (see appendix) to describe the dynamics of the DS qubit density matrix $\rho_{q}$ after taking trace over the phonon's degrees of freedom \cite{Agarwal1974,Lehmberg1970,Ficek2002}
\begin{eqnarray}
\frac{\partial\rho_{q}(t)}{\partial t}  &=& -\frac{i}{\hbar} \left[H_{q},\rho_{q}(t)\right] -\sum^{2}_{i\neq j}\eta_{ij}\left[\sigma_{+}^{i}\sigma_{-}^{j},\rho_{q}(t)\right]\nonumber \\&+&  \sum^{2}_{ij=1}\Gamma_{ij}\left[\sigma_{-}^{j}\rho_{q}(t)\sigma_{+}^{i}-\frac{1}{2} \lbrace \sigma_{+}^{i}\sigma_{-}^{j},\rho_{q}(t) \rbrace \right], \nonumber
 \label{master eq.}                                    
\end{eqnarray}
where
\begin{eqnarray}
\Gamma_{ij} &=& 2L\int_{0}^{\infty}dk g^{(i)}_{k}g_{k}^{(j)*} \delta(\omega_{k}-\omega_{0}) \nonumber \\
\eta_{ij} &=&\frac{L}{2\pi}\wp\int_{0}^{\infty}dk g^{(i)}_{k}g_{k}^{(j)*}\frac{1}{\left(\omega _{k}-\omega _{0}\right)},\label{parameters}
\end{eqnarray}
with $\wp$ standing for the principal value of the integral.
\begin{figure}[t!]
\includegraphics[scale=0.9]{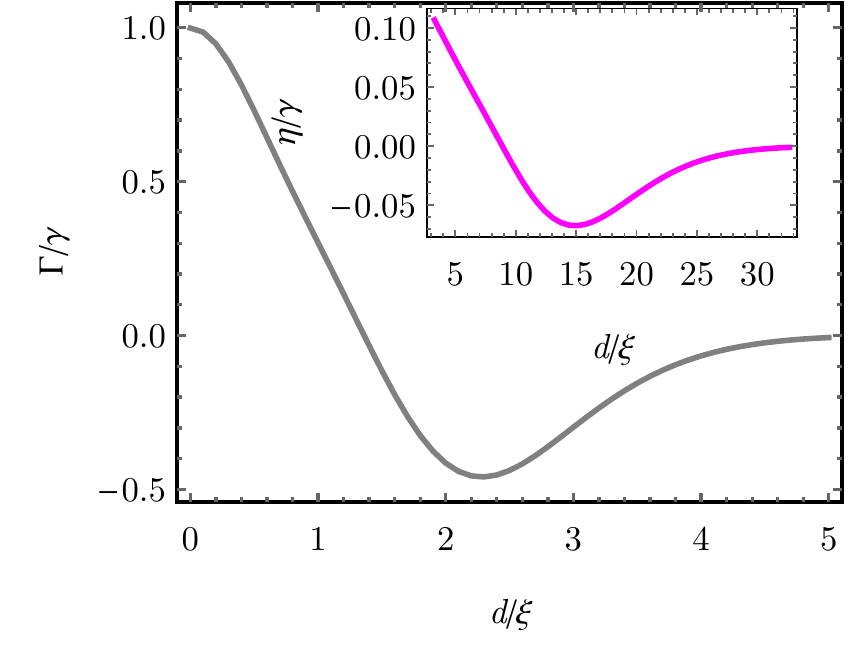}
\caption{(color online) Collective damping $\Gamma$ and qubit-qubit interaction parameter $\eta$ (inset) as a function of interatomic distance $d$. We have choosen $\nu=0.75$, for which DS qubit is well defined.}
\label{fig_damping}
\end{figure}
For $i=j$, $\Gamma_{ii}\equiv \gamma$ denotes the spontaneous emission rate of the qubits due to the Bogoliubov excitations (phonons); for $i\neq j$, $\Gamma_{ij}\equiv \Gamma$ is the collective damping resulting from the mutual exchange of phonons. The coherent term $\eta_{ij}\equiv \eta$ represents the interaction between DS qubits. Both the coherent and incoherent terms are dependent on the distance $d$ between DS qubits. Fig. (\ref{fig_damping}) depicts the dependence of the collective damping  $\Gamma$ and the coherent interaction $\eta$ as a function of the DS qubit distance $d$. For large separations, i.e. $d \gg\xi$, both parameters are very small (i.e. $\Gamma = \eta \approx 0$).\par

The main concern of the present work is the study of the time evolution of the entanglement. To study this, the most adequate sates are the collective, or the so-called Dicke, states \cite{Dicke1954}
\begin{eqnarray}
\left\vert g\right\rangle &=& \left\vert g_{1},g_{2}\right\rangle , \nonumber \\
\left\vert \pm\right\rangle &=& \left({\left\vert e_{1},g_{2}\right\rangle \pm \left\vert g_{1},e_{2}\right\rangle}\right)/{\sqrt{2}},  \nonumber \\
 \left\vert e\right\rangle &=& \left\vert e_{1},e_{2}\right\rangle ,\label{Collective states}
\end{eqnarray}
as schematically represented in Fig. (\ref{fig_Collective States}). The density matrix of Eq. (\ref{Collective states}) becomes
\begin{equation}
\rho =\left(                                                   
\begin{array}{cccc}                           
\rho _{ee} & \rho _{e+} & \rho _{e-} & \rho _{eg} \\                                  
\rho _{+e} & \rho _{++} & \rho _{+-} & \rho _{+g} \\                                  
\rho _{-e} & \rho _{-+} & \rho _{--} & \rho _{-g} \\                                
\rho _{ge} & \rho _{g+} & \rho _{g-} & \rho _{gg}
\end{array}%
\right) ,  \label{Dens. Mat}
\end{equation}
where $\rho_{ij} =\left\vert \psi_{i} \right\rangle \left\langle \psi_{j} \right\vert $. The elements of the above density matrix can be determined by using Eq. (\ref{master eq.}),
\begin{eqnarray}
\rho_{ee}(t)&=& e^{-2\gamma t}\rho_{ee}(0)\nonumber \\
\rho_{++}(t)&=&e^{-\left(\gamma+\Gamma\right) t}\rho_{++}(0) \nonumber \\ &+& \frac{\left(\gamma+\Gamma\right)}{\left(\gamma-\Gamma\right)}\left(e^{-\left(\gamma+\Gamma\right) t}-e^{-2\gamma t}\right)\rho_{ee}(0)\nonumber \\
\rho_{--}(t)&=&e^{-\left(\gamma-\Gamma\right) t}\rho_{--}(0)\nonumber \\ &+& \frac{\left(\gamma-\Gamma\right)}{\left(\gamma+\Gamma\right)}\left(e^{-\left(\gamma-\Gamma\right) t}-e^{-2\gamma t}\right)\rho_{ee}(0)\nonumber \\
\rho_{eg}(t)&=& e^{-\gamma t}\rho_{eg}(0)\nonumber \\
\rho_{+-}(t)&=& e^{-\left(\gamma-2i\eta\right) t}\rho_{+-}(0) \nonumber \\
\rho_{e+}(t)&=& e^{-\frac{1}{2}\left(3\gamma +\Gamma-2i\eta\right) t}\rho_{e+}(0)\nonumber \\
\rho_{e-}(t)&=& e^{\frac{1}{2}\left(3\gamma +\Gamma-2i\eta\right) t}\rho_{e-}(0)\nonumber \\
\rho_{g+}(t)&=& e^{-\frac{1}{2}\left(\gamma +\Gamma-2i\eta\right) t}\rho_{g+}(0)\nonumber \\ &+&\frac{\left(\gamma+\Gamma\right)}{\left(\gamma+2i\eta\right)}2e^{-\frac{1}{2}\left(2\gamma +\Gamma\right) t} \sinh\left(\frac{t}{2}\left(\gamma+2i\eta\right)\right)\rho_{+e}(0) \nonumber \\
\rho_{g-}(t)&=& e^{-\frac{1}{2}\left(\gamma -\Gamma+2i\eta\right) t}\rho_{g-}(0)\nonumber \\ &-&\frac{\left(\gamma-\Gamma\right)}{\left(\gamma-2i\eta\right)}2e^{-\frac{1}{2}\left(2\gamma -\Gamma\right) t} \sinh\left(\frac{t}{2}\left(\gamma-2i\eta\right)\right)\rho_{-e}(0) \nonumber \\
\label{den. mat ele.} 
\end{eqnarray}
with $\rho_{jk}=\rho^{*}_{kj}$ and $\rho_{gg}=1-\rho_{ee}-\rho_{++}-\rho_{--}$. Eq. (\ref{den. mat ele.}) depicts that all transition rates to and from
the state $\rho_{++}$ are equal to $\gamma+\Gamma$ while from state $\rho_{--}$ are $\gamma-\Gamma$. Therefore, the state $\rho_{++}$ decays with an enhanced
(superradiant) rate and $\rho_{--}$ with a reduced (subradiant) rate.
\section{Measurement of Entanglement}
The amount of entaglement can be determined by using the Wootter's concurrence \cite{Wootters1998}
\begin{eqnarray*}
C(t)=\max\left(0, \sqrt{\varepsilon}_{1}-\sqrt{\varepsilon}_{2}-\sqrt{\varepsilon}_{3}-\sqrt{\varepsilon}_{4}\right),
\end{eqnarray*}
where the $\varepsilon_{i}'s$ are the eigenvalues in decreasing order of magnitude of the matrix
\begin{eqnarray*}
\zeta=\rho\sigma_{y}\otimes\sigma_{y}\rho^{\ast}\sigma_{y}\otimes\sigma_{y}.
\end{eqnarray*}
Here, $\rho^{\ast}$ represents the complex conjugate of $\rho$ and $\sigma_{y}$ is the Pauli matrix. Depending on the initial state, concurrence can reach a value equal to zero assymptotically or at some finite time. It is interesting to observe that locally equivalent initial states with the same concurrence can disentangle at different times, depending on the parameters $\Gamma$ and $\eta$. In what follows, we investigate this aspect. 

\subsection{Entangled State}

\begin{figure}[t!]
\includegraphics[scale=0.6]{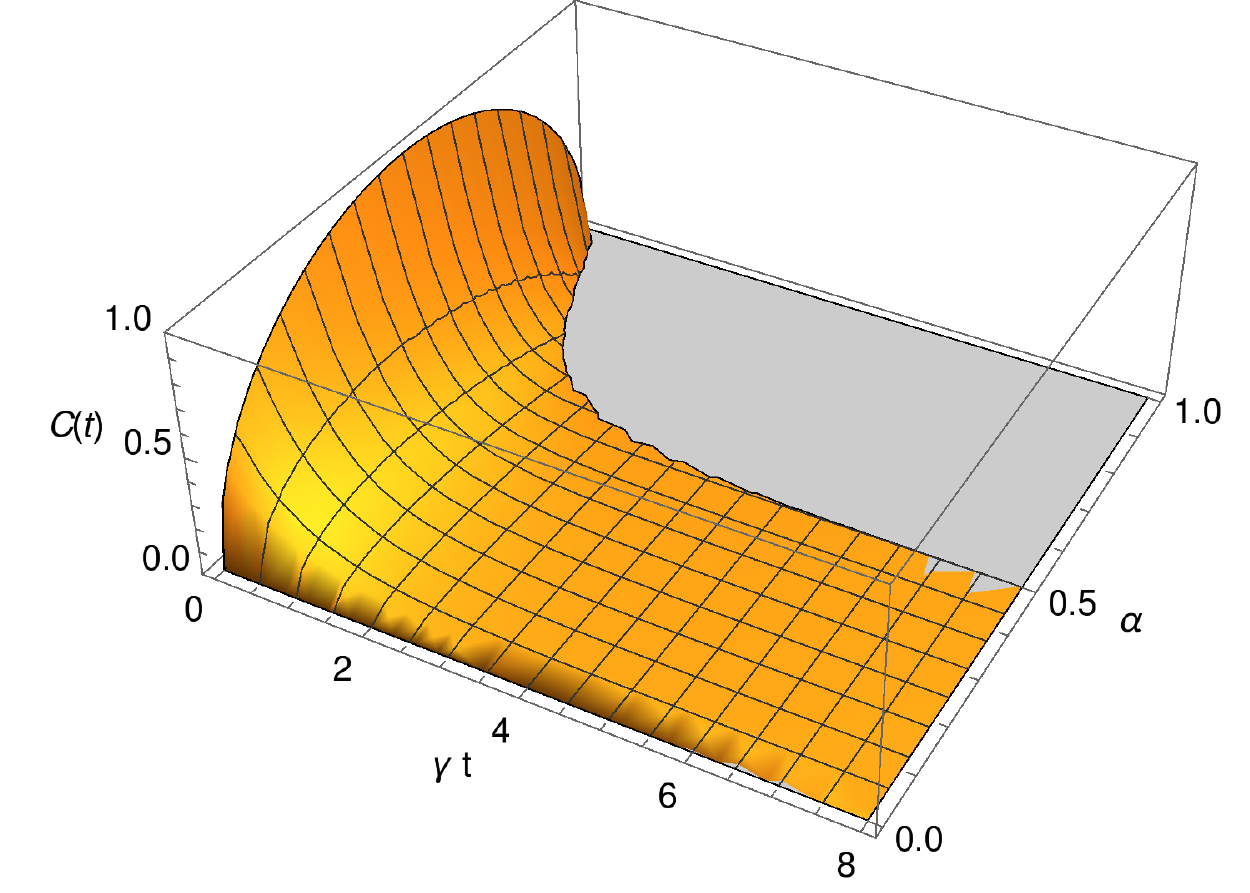}
\caption{(color online)  Time evolution of concurrence $C(t)$  for for initial an entangled state $\left\vert \Psi\right\rangle$ in non interacting DS qubit system. }
\label{fig_ind. qubits}
\end{figure}

Let assume that, initially, both or neither of the DS qubits are excited, i.e., the qubits are choosed to be prepared initially in an entangled state
\begin{eqnarray}
\left\vert \Psi\right\rangle=\sqrt{1-\alpha}\left\vert g \right\rangle + \sqrt{\alpha}\left\vert e\right\rangle , \label{initial state}
\end{eqnarray}
with $0\leq \alpha \leq 1$. Therefore, the density matrix Eq. (\ref{Dens. Mat}) becomes,
\[\rho =\left(                                                   
\begin{array}{cccc}                           
\rho _{ee} & 0 & 0 & \rho _{eg} \\                                  
0 & \rho _{++} & 0 & 0 \\                                  
0 & 0 & \rho _{--} & 0 \\                                
\rho _{ge} & 0 & 0 & \rho _{gg}%
\label{Fin. Den. Mat.}                                
\end{array}%
\right) . \]
The eigenvalues of the respective matrix $\zeta$ are thus given by
\begin{eqnarray*}
\sqrt{\varepsilon_{1,2}}&=&\sqrt{\rho_{ee}(t)\rho_{gg}(t)} \pm \vert \rho_{ge}(t) \vert \nonumber \\
\sqrt{\varepsilon_{3,4}}&=&\frac{1}{2}\left[ \rho_{++}(t)+\rho_{--}(t)\right] \pm  \frac{1}{2} \left[\rho_{++}(t)-\rho_{--}(t)\right].
\end{eqnarray*}
It is easy to verify that, depending on the largest eigenvalue (either $\sqrt{\varepsilon_{1}}$ or $\sqrt{\varepsilon_{3}}$), the concurrence $C(t)$ can be defined in two alternative ways, i.e.,
\begin{eqnarray}
C_{1}(t)&=&2\vert \rho_{ge}(t) \vert - \left[\rho_{++}(t)+\rho_{--}(t)\right] \label{c1}  \\
C_{2}(t)&=&\vert \rho_{++}(t)-\rho_{--}(t)\vert - 2\sqrt{\rho_{ee}(t)\rho_{gg}(t)}, \label{c2} 
\end{eqnarray}
where $C_{1}(t)$  measures the entanglement produced by the state of Eq. (\ref{initial state}) with the necessary condition $ \rho_{eg}(t) \neq 0$, while $C_{2}(t)$ provides the entanglement formed by the states $\left\vert \pm \right\rangle$. Notice that the positiveness of $C_{2}(t)$ is guaranteed if the latter states are not equally populated. 
At $t=0$, the system is entangled by the amount $C_{1}(0)=2\sqrt{\alpha(1-\alpha)}$. By inspecting Eqs. (\ref{den. mat ele.}), (\ref{c1}) and (\ref{c2}), it is possible to observe that DS qubits radiating independently ($\Gamma=0$) can not be entangled by following the criterion $C_{2}(t)$ because $\rho _{++}(t)=\rho _{--}(t)$. Thus, the time for the entanglement death due to spontaneous emission can be found via the condition $C_{1}(t)=0$, which provides
\begin{eqnarray}
t_{\rm death}=\frac{1}{\gamma}\rm ln\left(\frac{\alpha}{\alpha - \sqrt{\alpha(1-\alpha)}}\right). \label{death time} 
\end{eqnarray}
It is also pertinent to mention here that the condition for a finite-time disentanglement for independent DS qubits is $\alpha>1/2$ (see Fig. (\ref{fig_ind. qubits})). \par
\begin{figure}[t!]
\includegraphics[scale=0.6]{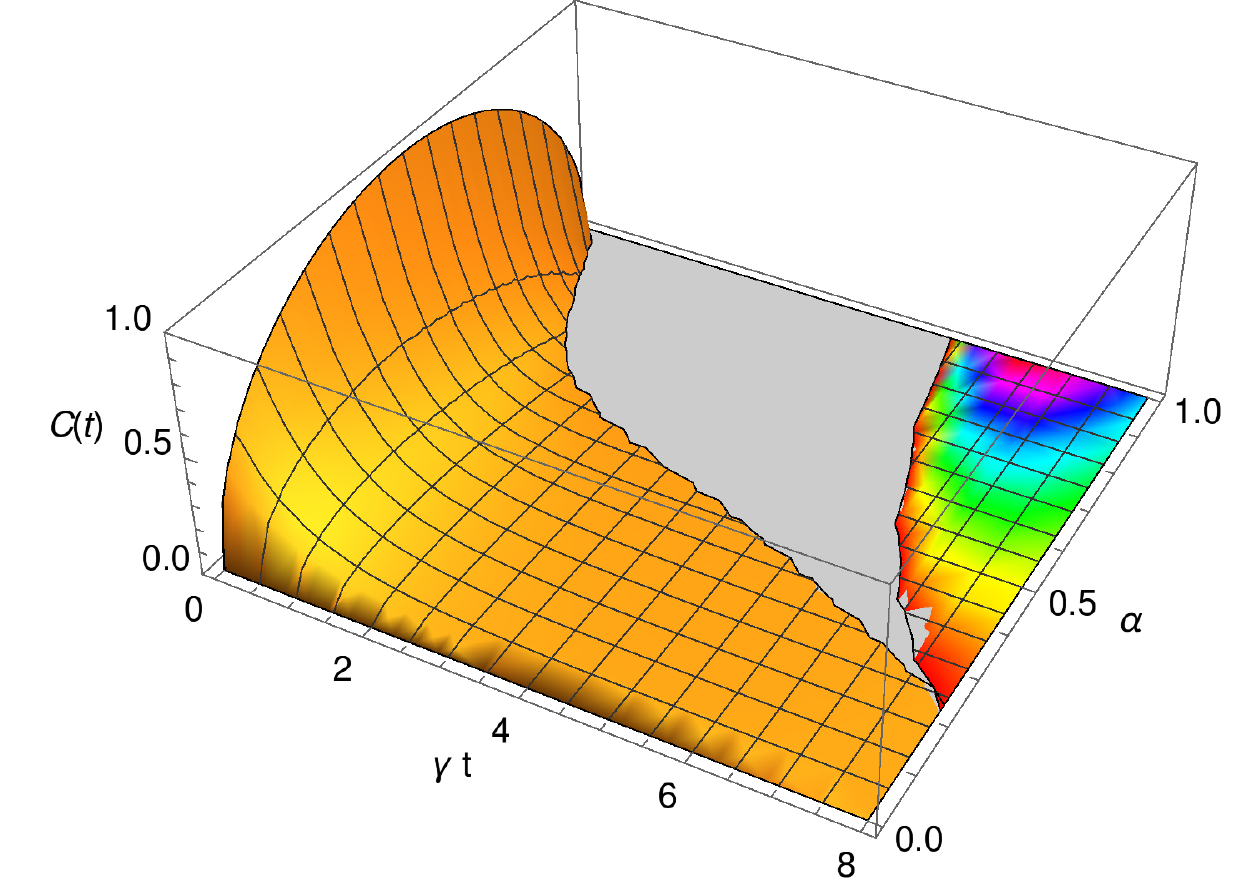}
\caption{(color online) Time evolution of concurrence $C(t)$  for initial an entangled state $\left\vert \Psi\right\rangle$ at distance $d=6\xi/5$.}
\label{fig_con.}
\end{figure}
\begin{figure}[t!]
\includegraphics[scale=0.5]{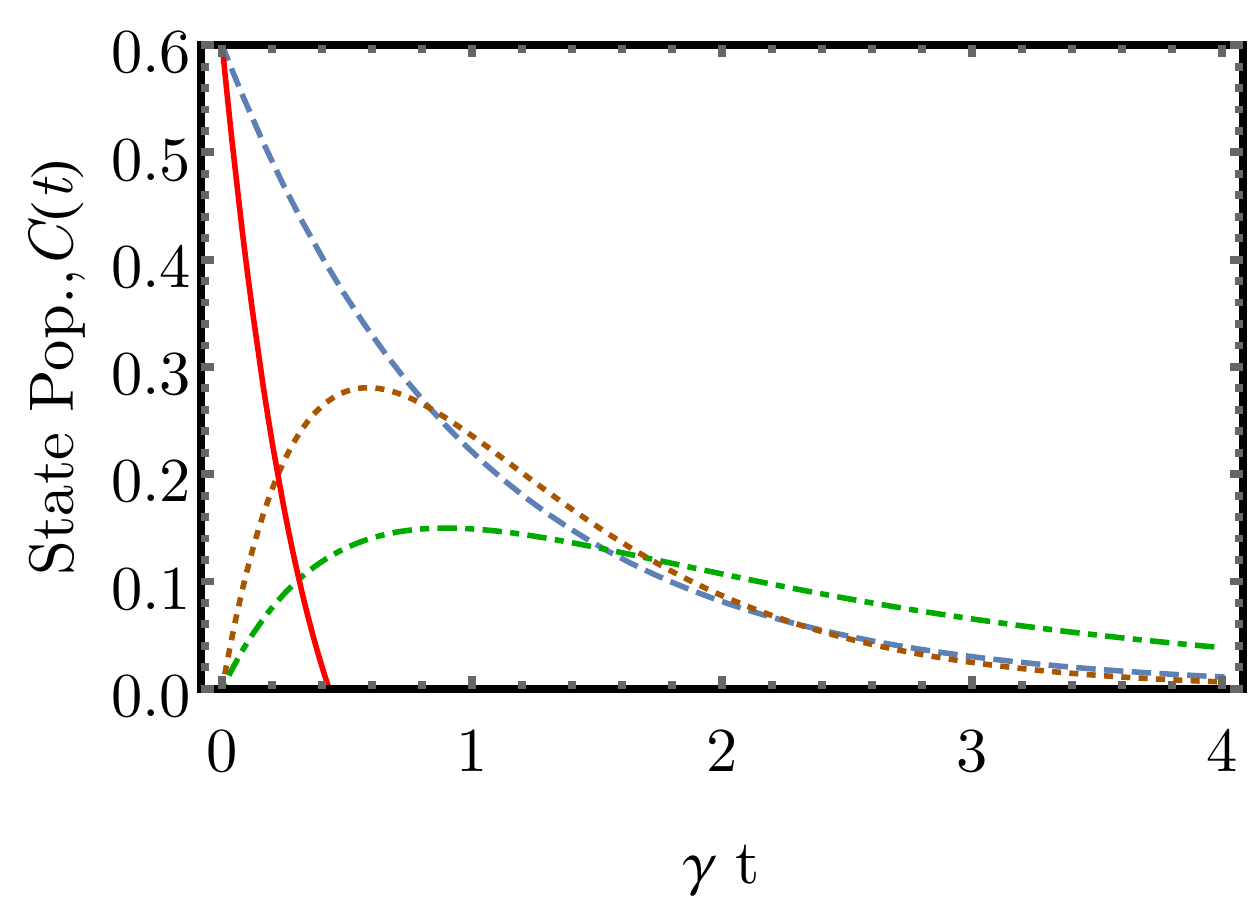}
\caption{(color online) State population (dashed curve for $\rho_{ge}(t)$, dotted-dashed for $\rho_{++}(t)$ and dotted for $\rho_{--}(t)$) and concurrence $C_{1}(t)$ (solid curve) at $ d \simeq 5\xi/2$. }
\label{fig_pop.}
\end{figure}

The situation changes when we allow the qubits to interact. In this case, the entanglement death is followed by its revival at a larger time, $t_{\rm revival}$. Fig. (\ref{fig_con.}) depicts the time evolution of the concurrence for the collective interactive system. It is shown that the entanglement dies as consequence of the spontaneous emission, but revives after a time $t_{\rm revival}\simeq 8/\gamma$, for $\alpha \simeq 1/4$ and qubits placed at a distance $d = 6\xi/5$. After a careful inspection of Eq. (\ref{den. mat ele.}), it is observed that the concurrence $C_{1}(t)<0$ at long times. Therefore, finite time ($t_{\rm revival}$) entanglement is determined by following $C_{2}(t)$ which yields. 
\begin{eqnarray}
t_{\rm revival}=\frac{2}{3\Gamma}\rm ln\left(\frac{4\gamma}{\sqrt{\alpha}(\gamma-\Gamma)}\right). \label{rev.2} 
\end{eqnarray}
Moreover, It can be analyzed from Fig. (\ref{fig_pop.}) that entanglement vanishes around the time at which $\rho_{--}(t)$ is maximally populated, and that it is not undergo any revival. The latter is due to the impartiality between the term  $\rho_{eg}(t)$ and $\rho_{--}(t)$. In other words, $\rho_{eg}(t)$ and $\rho_{--}(t)$ go to almost zero at long times while the population of $\rho_{++}(t)$ accumulates on the time scale $t=1/(\gamma+\Gamma)$ which is sufficiently large at $\Gamma \approx -0.5\gamma$. The collective damping $\Gamma \approx 0$ at farthest distance $d \simeq 5\xi$ for which DS qubits act like an independent qubits (fig. (\ref{fig_ind. qubits})).

\subsection{Mixed State}

\begin{figure}[t!]
\includegraphics[scale=0.6]{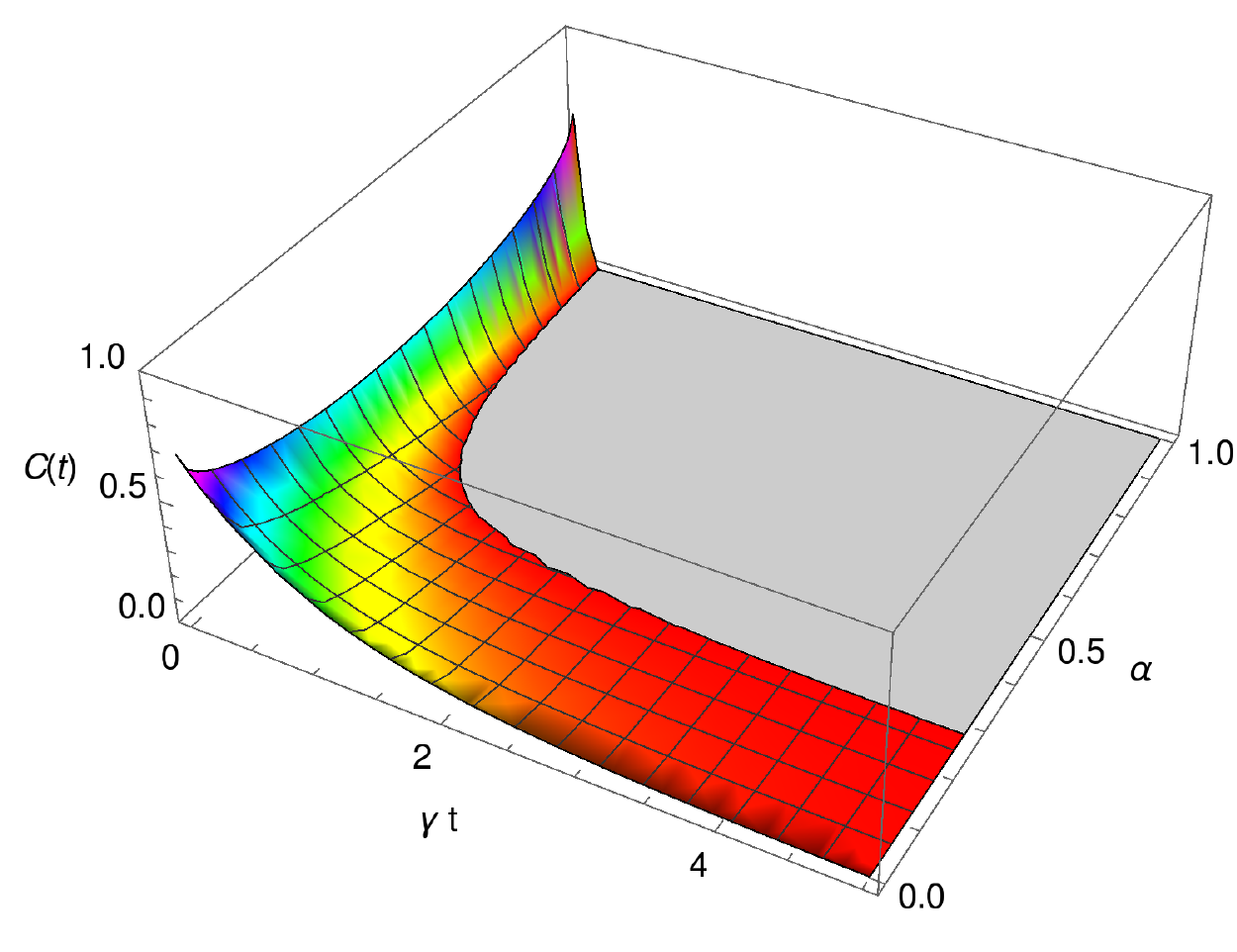}
\caption{(color online) Time evolution of concurrence $C(t)$ for a mixed initial state in non-interacting DS qubit system.}
\label{mixed state (ind)}
\end{figure}

\begin{figure}[t!]
\includegraphics[scale=0.6]{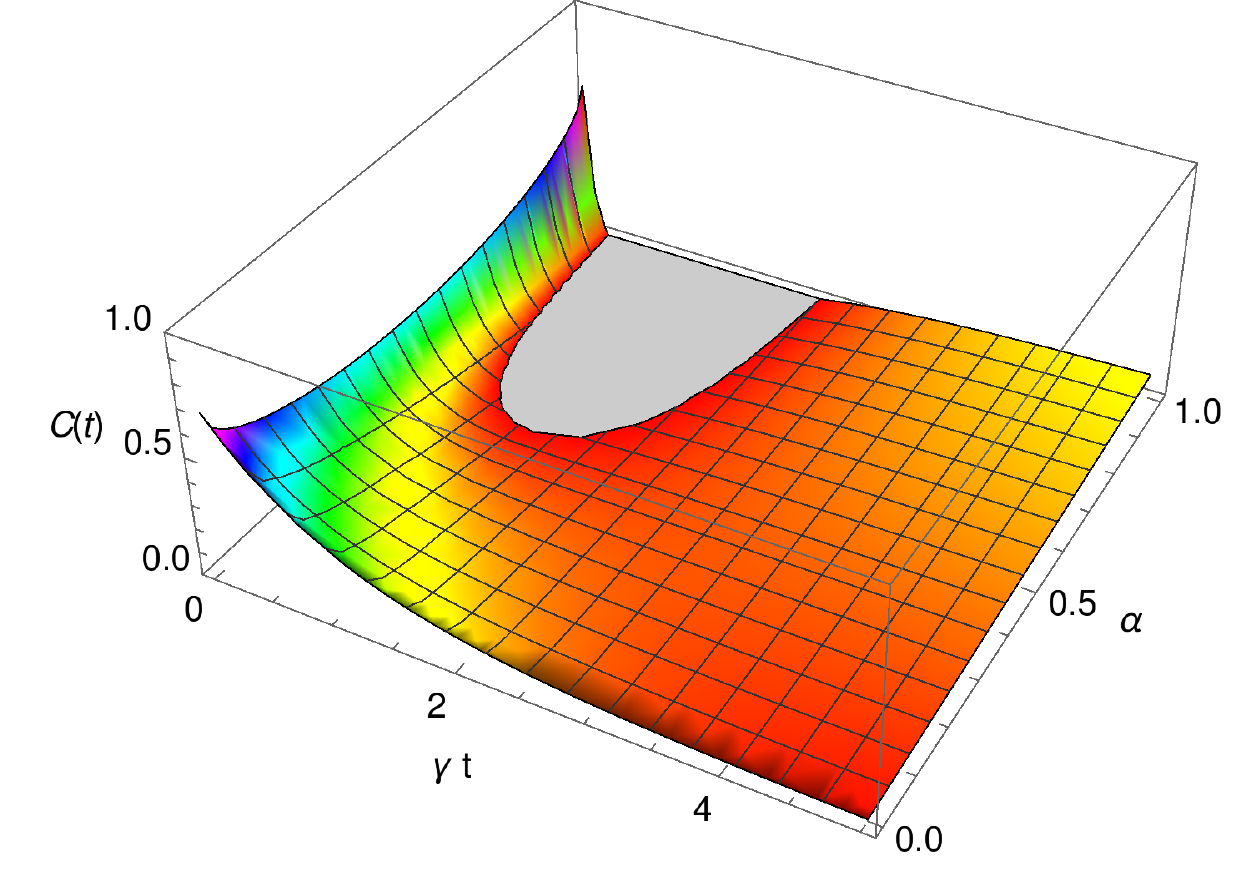}
\caption{(color online) Time evolution of concurrence $C(t)$ for a mixed initial state at distance $d\approx 4\xi$ between DS qubits.}
\label{mixed state}
\end{figure}

We now consider a two-qubit system to be initially prepared in a diagonal basis of the collective states. Therefore, the initial density matrix has the form
\begin{equation}
\rho(0) =\frac{1}{3}\left(                                                   
\begin{array}{cccc}                           
\alpha & 0 & 0 & 0 \\                                  
0 & 2 & 0 & 0 \\                                  
0 & 0 & 0 & 0 \\                                
0 & 0 & 0 & 1-\alpha%
\label{Fin. Den. Mat.}                                
\end{array}%
\right) , 
\end{equation}
The initial concurrence is determined by $C_{2}(0)=2\left(1-\sqrt{\alpha(1-\alpha)}\right)/3$, and the sudden-death time for independent DS qubits can be described by using criterion $C_{2}(t)=0$, which provides
\begin{eqnarray}
t_{\rm death}=\frac{1}{\gamma}\rm ln\left(\frac{\alpha}{\sqrt{3\alpha^{2}+5\alpha)}-(1+\alpha)}\right). \label{mixed states death time} 
\end{eqnarray}
It is obvious from Eq. (\ref{mixed states death time}) that the ESD is possible only for $\alpha \gtrsim 1/3$ (see Fig. (\ref{mixed state (ind)})). The time evolution of the concurrence for interacting qubits is depicted in Fig. (\ref{mixed state}). It is observed that the entanglement first decay and then revives for $\alpha\gtrsim1/2$ at $d\approx 4\xi$. ESD happens at $t_{\rm death} \sim 0.75/\gamma$, while entanglement revival is obtained at $t_{\rm revival}\sim 1.4/\gamma  $. 

\subsection{Experimental estimates} 

We consider a quasi 1D BEC of $^{85}$Rb with a chemical potential ($\mu$) of a few kHz. This yields a qubit gap frequency $\omega _{0}/2\pi\sim 0.5$ kHz, the spontaneous decay rate $\gamma/2\pi \sim 29$ Hz and a collective decay $\Gamma/2\pi \sim 6$ Hz, at $d \sim 6\xi/5\sim 1$ $\mu$m. These rates validate a posteriori the RWA and Markovian approximations. Therefore, the sudden death-time for the maximally entangled state is $t_{\rm death} \sim 19$ ms and the revival time is $t_{\rm revival} \sim 35$ ms, where the period $ \Delta t \sim 16$ ms is the dark period, i.e. the time interval during which $C(t)=0$. For large separations, ESD occurs at $t_{\rm death} \sim 2$ ms due to the balanced population of $\rho_{eg}$ and $\rho_{--}$ whereas, for the mixed initial state, the dark period occurs for $\sim 3.6$ ms. Prolonging entanglement is essential for practical realization of quantum information and computation protocols based on entanglement. Therefore, the dark period of entanglement can be delayed or averted by carrying out local unitary operation on qubits \citep{Rau2008, Singh2017}.

\section{Summary and Discussion}

To summarize, we investigate the finite-time disentanglement, or the entanglement sudden death, between two dark-soliton qubits produced in a quasi one-dimensional BEC. We derive the master equation and extracted the time evolution of the relevant density matrix elements. The Wooter's concurrence is used as a measure of entanglement and we show the collapse and revival behavior, depending on the collective damping and on the initial state. For initial an entangled state, the concurrence can not be revived at large distances in the range of $2-5$ $\mu$m due to the impartial behavior of the populated states, while it revives for a mixed state. Therefore, it can be concluded that the collective behavior of the dark soliton qubits reveals the dependence of entanglement evolution on the interatomic distance and it becomes quite different from that of independent dark soliton qubits.


\appendix 
\section{Derivation of the master equation}

We begin by writing the total Hamiltonian by spanning the Hilbert space as
\begin{eqnarray}
H(t)= H_{q}\otimes I + I \otimes H_{p} + H_{\rm int}(t),
\end{eqnarray}
The key ingredient for the application of the Born-Markov approximation is the assumption that $H_{q} \otimes H_{p}$ is small compared to the remaining terms, so that a perturbative treatment of the interaction is possible. To make this more explicit, we take $H_{0}=H_{q} + H_{p}$ and move into the interaction picture
\begin{equation}
H_{I}(t)=e^{i H_{0} t}H_{I}e^{-i H_{0} t}.
\end{equation}
We make the Born approximation and assume that the density operator factorizes at all times as
\begin{eqnarray}
\rho(t)\cong \rho_{q}(t) \otimes \rho_{p},
\end{eqnarray}
where the reservoir density operator is assumed to be time independent i.e., $\rho_{p}=\rho_{p}(0)$. Therefore, within the interaction picture, the density operator evolves according to
\begin{equation}
\frac{d\rho (t)}{dt}=-{\imath}[H_{I}(t),\rho(t)], 
\label{1}
\end{equation}
and the respective formal integral solution is given by
\begin{equation}
\rho(t)=\rho(0)-{\imath}\int_{0}^{t}[H_{I}(s),\rho(s)]ds. \label{2}
\end{equation}
By plugging Eq. (\ref{2}) into Eq. (\ref{1}), and by taking the partial trace over reservoir degrees of freedom (phonons), we obtain
\begin{equation}
\frac{d\rho_{q}(t)}{dt}=-\int_{0}^{t}ds tr_{P}[H_{I}(t),[H_{I}(s),\rho_{q}(s)\rho_{p}]],\label{Redfield Eq.}
\end{equation}
This equation is called the Redfield equation where the term $ \mathrm{Tr}_{P}[H_{I}(t),\rho(0)]$ is disregarded. Further, we make use of the Markov approximation to put Eq. (\ref{Redfield Eq.}) into a more amenable form. This assures that the behaviour of $\rho_{q}(t)$ is local in time. This master equation is still depends on the choice of the initial state. However, making the substitution $s \rightarrow t-\tau$  and letting the upper integration limit to go to infinity, we obtain
\begin{equation}
\frac{d\rho_{q}(t)}{dt}=-\int_{0}^{\infty}d\tau tr_{P}[H_{I}(t),[H_{I}(t-\tau),\rho_{q}(t) \rho_{p}]]\label{Born Mark}
\end{equation}%

The latter is the Born-Markov master equation. To gain a little more insight into the structure of Eq. (\ref{Born Mark}), it is useful to be more specific about the form of the interaction picture Hamiltonian 
\begin{equation}
 H_{I}(t)= S^{\dagger}B+ SB^{\dagger},
\end{equation}
where $S^{\dagger}(t)=\sum_{i=1}^{2}\sigma^{(i)} _{+} = e^{{\imath}\omega _{0}t}S^{\dagger}$, $B(t)=\sum_{k} g(k)e^{{-\imath}\omega _{k}t}b$ and vice versa. Here, we use the identity
\begin{equation}
e^{{\alpha}A}Se^{-{\alpha}A}= S+\alpha[A,S]+\frac{\alpha^{2}}{2!} [A,[A,S]]+\cdots
\end{equation}
Moreover, we invoke the cyclic property of trace to write the revervoir correlation function as $\mathrm{Tr}_P\left(b_{k}b^{\dagger}_{q}\rho_{P}\right)=\delta_{k,q}$. As such, the Born-Markov master Eq. (\ref{Born Mark}) can be finally rewritten as 
\begin{eqnarray}
\frac{d\rho_{q}(t)}{dt}&=& -\gamma\left[ \rho_{q}(t) S^{\dagger}S-S\rho_{q}(t)S^{\dagger} \right]\nonumber \\ &-& \gamma\left[\rho_{q}(t)SS^{\dagger}-S\rho_{q}(t)S^{\dagger} \right] + {\rm h.c.},
\label{master eq. 1}
\end{eqnarray}
where 
\begin{eqnarray*}
\gamma  = \sum_{k} g(k)g(k)^{*}\int_{0}^{t}d\tau e^{{-\imath}\left(\omega _{k}-\omega _{0}\right)(t-\tau)}.
\end{eqnarray*}
The sum over the phonon $k-$ modes can be computed by taking the continuum limit
\begin{equation}
\sum_{k} \rightarrow \int_{0}^{\infty} D(k)dk,
\end{equation}
where $D(k)= L/2 \pi$ is the density of states, $L$ is the size of the system, and 
\begin{equation}
\int_{0}^{t}d\tau e^{{-i}\left(\omega _{k}-\omega _{0}\right)(t-\tau)}=\left(\pi\delta(\omega_{k}-\omega_{0})-i\wp/{\left(\omega _{k}-\omega _{0}\right)}\right).
\end{equation}
Transforming Eq. (\ref{master eq. 1}) back in the Schr\"odinger picture, we finally obtain
\begin{eqnarray}
\frac{d\rho_{q}(t)}{dt}  &=& -\frac{i}{\hbar} \left[H_{q},\rho_{q}(t)\right]-\sum^{2}_{i\neq j}\eta_{ij}\left[\sigma_{+}^{i}\sigma_{-}^{j},\rho_{q}(t)\right] \nonumber \\
 &+&\sum^{2}_{ij=1}\Gamma_{ij}\left[\sigma_{-}^{j}\rho_{q}(t)\sigma_{+}^{i}\right. \nonumber\\ 
 &-& \left.\frac{1}{2} \lbrace \sigma_{+}^{i}\sigma_{-}^{j},\rho_{q}(t) \rbrace \right] 
 \label{final master eq.}
\end{eqnarray}
Eq. (\ref{final master eq.}) is the final form of the master equation used in this work.


\section*{Acknowledgements}

The authors acknowledge the support from DP-PMI programme and Funda\c{c}\~{a}o para a Ci\^{e}ncia e a Tecnologia (Portugal), namely through the scholarship number SFRH/PD/BD/113650/2015 and the grant number SFRH/BPD/110059/2015 and No. IF/00433/2015. E.V.C. acknowledges partial support from FCT-Portugal through Grant
No. UID/CTM/04540/2013.

\end{document}